\documentclass{PoS}

\title{Hercules X-1 - another 'first': long-term decay of the cyclotron line energy}

\ShortTitle{Cyclotron line in Her X-1}

\author{\speaker{R\"udiger Staubert}\thanks{highlighting the discovery of the "Long-term change
in the cyclotron line energy in Hercules X-1", by R. Staubert, D. Klochkov, K. Postnov, N. Shakura, 
J. Wilms, R. Rothschild, F. F\"urst and F. Harrison, 2014, A\&A \textbf{572}, A119 (arXiv/1410.3647)} \\
        Author affiliation: Institut f\"ur Astronomie und Astrophysik, Univ. T\"ubingen, Germany\\
        E-mail: \email{staubert@astro.uni-tuebingen.de}}

\abstract{Her~X-1 is one of the most remarkable members of the class of binary 
X-ray pulsars. It does not only show a large number of observable
features, but has repeatedly been the first object for which fundamental
discoveries were made: besides being second of the the first two discovered X-ray binary 
pulsars (after Cen~X-3), it was the first to show a super-orbital modulation (with
systematic variations of the shape of the pulse profiles), the 
first to reveal a cyclotron line in its spectrum and the first in which systematic variations
of the cyclotron line energy were detected, namely variations with pulse phase (by 
$\sim$ 25\%) and a positive correlation with X-ray luminosity ($\sim$ 5\% increase for a 
factor of two increase in luminosity). Now we have found another 'first': a long-term 
decrease of the pulse phase averaged cyclotron line energy E$_{\rm cyc}$ by 
$\sim$ 5\,keV in 20 years (from 1992 to 2012). At the time of the discovery of the cyclotron line
in 1976, its energy was $\sim$ 35\,keV, remeasured around a similar energy by various
instruments until 1990. Between 1990 and 1994 a jump upwards beyond 41\,keV occurred.
Our current result of a slow decay thereafter - we are now back at $\sim$ 37\,keV - is based 
on repeated observations of Her~X-1 by all those X-ray observatories capable of measuring 
clearly beyond the cyclotron line energy.

With respect to the physical cause of the discovered effect, we suggest it to be connected 
to a geometric displacement of the cyclotron resonant scattering region in the polar magnetic field or 
due to to a true physical change in the field configuration at the accretion mound or column by the 
continued accretion. The variation with time might be due to a non-perfect equilibrium between the
rate at which material is accreted and the rate at which material is lost at the base of the accretion 
mound, allowing for a variation of the configuration of the accretion mound (height, total mass, field
distribution). We also speculate that the upward jump in E$_{\rm cyc}$ observed around 1991 may
have been due to a relatively fast event in which the polar magnetic field rearranged itself 
after releasing part of the accumulated material to larger areas of the neutron star surface.
We do believe that we see the signature of a local change in the field configuration, rather than
a decay of the global magnetic field, since the observed timescale of a few decades is very short. }

\FullConference{10th INTEGRAL Workshop: "A Synergistic View of the High Energy Sky" - Integral2014,\\
		15-19 September 2014 \\
		Annapolis, MD, USA }

\begin{document}

\section{Introduction}
\label{sec:introduction}

Her~X-1 is singled out among all known binary X-ray pulsars for the richness of observable
features and by the large amount of observational data collected since its discovery by UHURU
in 1971. Her~X-1 consists of a neutron star accreting from a 2.2 
$M_{\odot}$ optical companion through Roche lobe overflow and an accretion disk. The X-ray
source shows periodic variability on several different timescales: the 1.24\,s spin period of the 
neutron star, the 1.7-day binary period, the 35-day on-off period, and the 1.65-day period of the 
pre-eclipse dips. Some of the observational features are due to the high inclination of the
orbital plane.

Her~X-1 was detected by UHURU in 1971 and identified as the second accreting binary pulsar
(Tananbaum et al. 1972), after Cen~X-3 and by the same chain of arguments, that is by finding
regular eclipses and a sinusoidal modulation of the pulse arrival times, which suggests a near
circular orbit with a well defined orbital period. Today we know $\sim$ 200 X-ray binary pulsars 
(XRBP, Orlandini 2014).

In further UHURU observations the 35-day on-off modulation and the dips were detected 
(Giacconi et al. 1972). The two types of dips (\textsl{pre-eclipse dips} and \textsl{anomalous dips}) 
are very interesting features and their detailed (spectral and timing) analysis has contributed much
to the understanding of the mass transfer process in this binary (Gerend \& Boynton 1976,
Crosa \& Boynton 1980). They are thought to be due to absorption by the gas stream which
transports the material from the optical companion to the accretion disk, whenever the line of
sight to the neutron star is intersected. The dip timing is therefore intimately connected to both
the 1.7\,d orbital period and to the 35\,d precession period of the accretion disk.

Our current understanding of the 35-day modulation is that it is due to the precession of a warped 
accretion disk. Because of the high inclination ($i>80^\circ$) of the binary we see the disk nearly 
edge-on. The precessing warped disk covers the central X-ray source during a substantial portion 
of the 35-day period. Furthermore, a hot X-ray heated accretion disk corona reduces the X-ray 
signal (energy independently) by Compton scattering whenever it intercepts our line of sight 
to the neutron star. As a result, the X-ray source is covered twice during a 35-day cycle.
Another 35\,d modulation is present in the systematic variation of the shape of the 1.24\,s
pulse profile. It has been suggested by Tr\"umper et al. (1986) that the reason for this
is free precession of the neutron star, leading to a systematic change in our viewing angle to
the X-ray emitting regions. Postnov et al. (2013) have successfully reproduced observed
X-ray pulse profiles in the 9-13\,keV range by a model assuming pencil beam emission from
a special combination of point-like and arc like regions around the magnetic poles. 
There is, however, no generally accepted model for the generation of the complex pulse
profiles of Her~X-1 (which can formally well be fitted by eight equidistantly spaced Gaussian profiles),
nor about the systematic variations with 35-day phase. Scott et al. (2000) assume that the
inner edge of the accretion disk plays an important role, needing, however, a rather small
magnetospheric radius, which is not consistent with other estimates (see discussion in
Staubert et al. 2013). An open question is, whether the accretion disk and the neutron star could 
precess in synchronization due to a closed loop physical feed-back (Staubert et al. 2009)
(for which there is independent evidence, e.g. by the correlation between the histories of the
35-day turn-ons and the pulse period evolution; Staubert et al. 2006).
Further analysis of the variations in pulse profiles (Staubert et al. 2013), has shown that the irregular histories 
of the turn-ons (which we clearly attribute to the behavior of the accretion disk) and of the variations 
in the pulse shape are identical, with synchronized variations even on short timescales ($\sim300$\,d), 
indicating again the importance of the accretion disk for the generation of the pulse profiles.
In any case, Her~X-1 is again the first, and so far the only object, in which we can study the
above discussed phenomena in detail. The other two binary pulsars showing clear super-orbital 
modulations, LMC~X-4 and SMC~X-1, are not nearly as coherent in their behavior as Her~X-1; 
also, they have not been observed as thoroughly.

It is believed that the X-ray spectrum emerges from the hot regions around the magnetic 
poles where the accreted material, which is channeled by the $\sim10^{12}$\,G magnetic field 
towards the surface of the NS, is decelerated and where the kinetic energy is converted 
to heat and finally X-ray radiation. The height of the accretion mound is thought to be between 
a few tens and a few hundred meters. If the magnetic and spin axes of the neutron star are 
not aligned, a terrestrial observer sees a flux modulated at the rotation frequency of the star.
The X-ray spectrum of Her~X-1 is characterized by a power law continuum 
with exponential cut-off and an apparent line-like feature. The continuum is
believed to be due to thermal bremsstrahlung radiation from the $\sim$10$^{8}$\,K
hot plasma modified by Comptonisation (Becker \& Wolff 2007, Becker et al. 2012).
The line feature was discovered in 1976 in a balloon observation (Tr\"umper et al. 1978).
This feature is now generally accepted as an absorption feature around
40\,{\rm keV} due to resonant scattering of photons off electrons on
quantized energy levels (Landau levels) in the Teragauss magnetic
field at the polar cap of the neutron star. The feature is therefore
often referred to as a \textsl{Cyclotron Resonant Scattering Feature} (CRSF).  
The energy spacing between the Landau levels is given by E$_{\rm cyc}$ =
${\rm \hbar}$eB/m$_{\rm e}$c = 11.6\,{\rm keV}\,B$_{12}$, where
B$_{12}$=B/10$^{12}$\,{\rm G}, providing a direct method of measuring
the magnetic field strength at the site of the emission of the X-ray spectrum. The 
observed line energy is subject to gravitational redshift $z$ at the location where 
the line is formed, such that the magnetic field may be estimated by
B$_{12}$ = (1+z)~E$_{\rm obs}$/11.6\,${\rm keV}$, with E$_{\rm obs}$ being
the observed cyclotron line energy.  The discovery of the
cyclotron feature in the spectrum of Her X-1 provided the first ever
direct measurement of the magnetic field strength of a neutron star,
in the sense that no other model assumptions are needed.  
Originally considered an exception, cyclotron features are now known to 
be rather common in accreting X-ray pulsars, with $\sim25$ binary pulsars 
now being confirmed cyclotron line sources. Several objects show 
multiple lines (up to four harmonics in 4U~0115+63). Reviews are given by 
e.g., Staubert (2003), Heindl et al. (2004), Terada et al (2007), Wilms (2012),
Caballero \& Wilms (2012); see also the list of cyclotron line sources by Orlandini (2014). 
Theoretical calculations of  cyclotron line spectra
have been performed either analytically (Ventura et al. 1979, Nagel 1981, Nishimura 2008)
or making use of Monte Carlo techniques (Araya \& Harding 1999, Araya-Gochez \& Harding 
2000, Sch\"onherr et al 2007).

Here we like to highlight the new result of the long-term decrease in the energy E$_{\rm cyc}$
of the CRSF in the pulse averaged X-ray spectrum of Her~X-1, which is found to co-exist with
the correlation between E$_{\rm cyc}$ and the X-ray luminosity detected earlier. 
In addition, we discuss the historical evolution since the discovery of the CRSF in 1976. 
With regard to the physics, we speculate about the reasons for both the long-term decrease 
and the earlier observed fast upward jump as being connected to changes in the configuration 
of the magnetic field at the site of the emission above the polar caps of the accreting neutron star. 
The details of the underlying observations and of the data analysis are presented in Staubert 
et al. (2014). 

\newpage

\section{Variation of the cyclotron line energy E$_{\rm cyc}$}
\vspace{-2mm}

Variability in the CRSF energy in Her X-1 is found with respect to: \\
- Phase of the 1.24\,s pulsation (here not discussed further, see Vasco et al. 2013). \\
- Phase of the 35\,d precessional period (here not discussed further, see Staubert et al. 2014). \\
- X-ray luminosity (here and Staubert et al. 2007, Klochkov et al. 2011, Staubert et al. 2014). \\
- Time: a true long-term decay (here and Staubert et al. 2014). 

\begin{figure}
\includegraphics[angle=90,width=14cm]{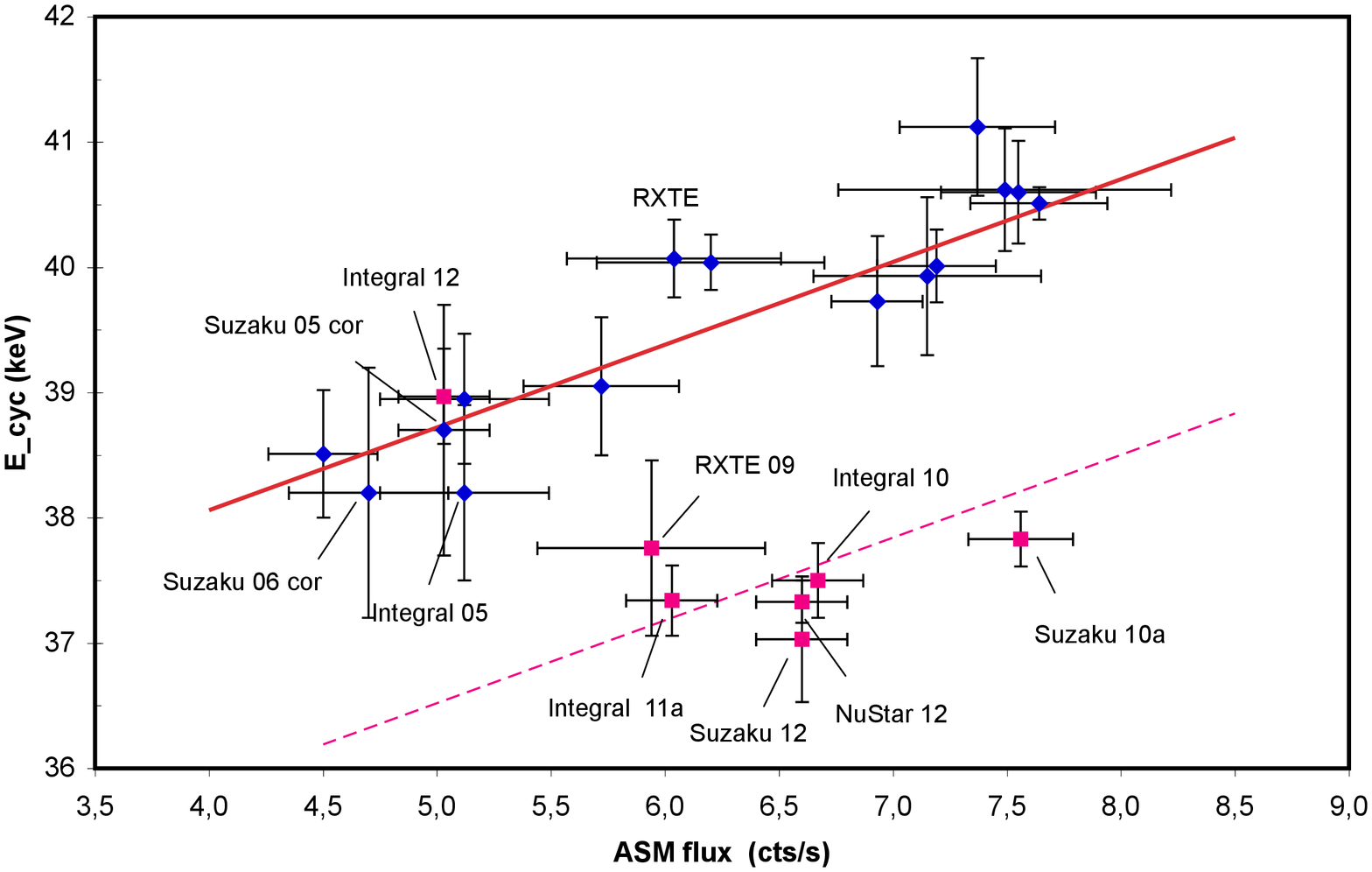}
\hfill
\vspace{-10mm}
\caption{The positive correlation between the cyclotron line energy and the maximum X-ray 
  flux of the corresponding 35-day cycle as measured by \textsl{RXTE/ASM} (see Fig.~2 of 
  Staubert et al. 2007) with eight added points: \textsl{INTEGRAL} 2005 
 (Klochkov et al. 2008), \textsl{Suzaku} of 2005 and 2006 (Enoto et al. 2008), 
  \textsl{RXTE} 2009, \textsl{INTEGRAL} 2010, \textsl{Suzaku} 2010 and 2012 and \textsl{NuSTAR} 2012.
  The \textsl{Suzaku} points of 2005/2006 have been corrected upward by 2.8\,keV, 
  to account for the difference arising because the Lorentzian profile was used in the analysis by  
 Enoto et al. (2008), while for all others the Gaussian profile was used. 
  The blue rhombs are values observed until 2006, the red dots are from after 2006.
  The solid red line is a linear fit to data until 2006 with the original slope of
  0.66\,keV/(ASM cts/s), as found by Staubert et al. (2007).
  The dotted red line is the best fit to the data after 2006 with the slope fixed to the same value.}
   \label{fig:correlation}
\end{figure}

\section{Variation of E$_{\rm cyc}$ with luminosity}
\label{sec:luminosity-dependence}
\vspace{-2mm}

For Her~X-1, the dependence of the centroid energy of the phase averaged cyclotron 
line on X-ray flux was discovered by Staubert et al. (2007) while analyzing a uniform
set of observations from \textsl{RXTE}. The original aim of the analysis at that time had 
been to investigate a possible decrease in the phase averaged cyclotron line energy with 
time during the first decade of \textsl{RXTE} observations. Instead, the dependence on 
X-ray flux was discovered and shown that the apparent decrease in the measured values of 
the line energy was largely an artifact due to this flux dependence. 
The correlation was found to be positive, that is the cyclotron line energy E$_{\rm cyc}$ 
increases with increasing X-ray luminosity $L_x$. 

Fig.~\ref{fig:correlation} reproduces the original correlation graph of Staubert et al. (2007)
with new data points added. The first three new data points (INTEGRAL~05 and Suzaku~05/06) 
fit very well into the previous data set (and do not change the formal correlation - see the solid red 
line), but most of the values from 2006-2012 are significantly lower. 
As we will show below, it is these data which clearly establish a decrease in the cyclotron line energy with time.
After 2006 the flux dependence is less obvious. However, the data points (except the one from \textsl{INTEGRAL} 2012)
are consistent with the originally measured slope (0.66 keV/ASM-cts/s) with generally lower E$_{\rm cyc}$ values. 
The dotted red line is a fit through the data after 2006 with the same slope as the solid red line.
We note that \textsl{flux} refers to the maximum \textsl{Main-On} flux as determined using the \textsl{RXTE}/ASM
and/or the \textsl{Swift}/BAT monitoring data (since 2012 from BAT only); the conversion is: 
(2-10\,keV ASM-cts/s) = 89 $\times$ (15-50\,keV BAT-cts~cm$^{-2}$~s$^{-1}$).
The \textsl{INTEGRAL} 2012 point does clearly not follow this behavior, as will be more obvious below. 
We have invested a considerable effort to check the calibration of the \textsl{INTEGRAL}/ISGRI detector 
(INTEGRAL Soft Gamma-Ray Imager) for the time of the observation and the 
data analysis procedure. The ISGRI response was closely examined by us for each of our Her X-1 observations. 
When necessary, the ARFs (Auxilliary Response Files) were checked (using the nearest Crab observations) and the 
energy scale was individually controled by making use of observed instrumental background lines 
with known energy. Finally, spectra were generated using data from SPI (Spectrometer onboard INTEGRAL): 
the resulting E$_{\rm cyc}$ values were always consistent with those of the ISGRI analysis.
Since we have found no errors, we keep this point in our data base, but will exclude it from some of the 
analysis discussed below.

\section{Variation of E$_{\rm cyc}$ with time - long-term variation}
\label{sec:secular}
\vspace{-2mm}

\begin{figure*}
\includegraphics[width=0.65\textwidth,angle=-90]{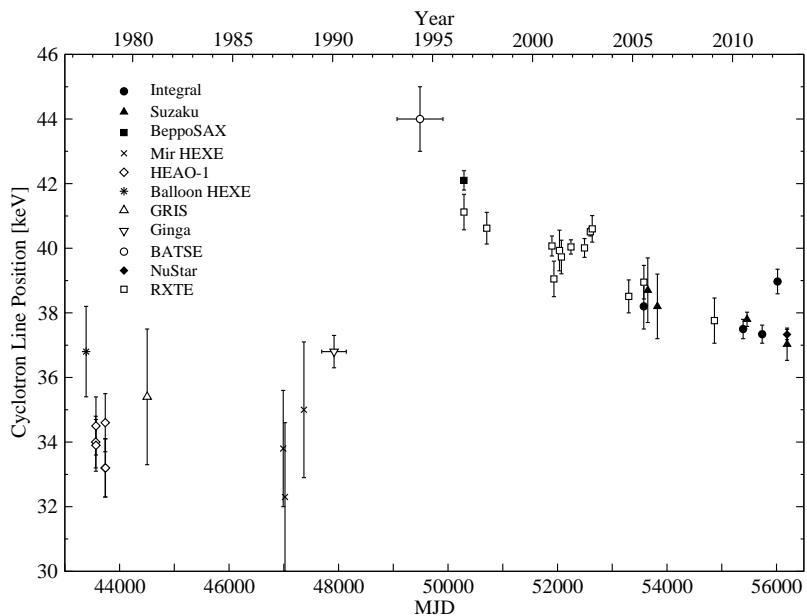}
\vspace{-4mm}
\caption{The centroid energy of the phase averaged cyclotron resonance
  line feature in Her~X-1 since its discovery. Data from before 1997
  were originally compiled by Gruber et al. 2001, where the original references can be found. 
The data after 1997 are from observations by \textsl{RXTE}, \textsl{INTEGRAL} 
(Klochkov et al. 2006, Staubert et al. 2007, Klochkov et al. 2008) and \textsl{Suzaku}
(Enoto et al. 2008), plus recent values as listed in Table~2 of Staubert et al. (2014).
Here only values measured at 35d phases $< 0.20$ are shown.
}
 \label{fig:history}
\end{figure*}

In Fig.~\ref{fig:history} (an update of Fig.~\ref{fig:correlation} of Staubert et al. 2007) we display 
observed values of the pulse phase averaged centroid cyclotron line energy as a 
function of time, covering the complete history of observations since the discovery of 
the line in 1976. We combine historical data, as taken from the compilation by 
Gruber et al. (2001) (their Tables~2 and~3) for the time before the \textsl{RXTE} era, 
published values from observations with \textsl{RXTE} and \textsl{INTEGRAL} 
(Klochkov et al. 2006, Staubert et al. 2007, Klochkov et al. 2008) and with
\textsl{Suzaku} (Enoto et al. 2008), as well as recent values (see Staubert et al. 2014).
For the analysis of the long-term variation of E$_{\rm cyc}$ we exclude values with 
35\,d phases $>$0.20 in order to avoid contamination due to a possible third variable, 
the 35\,d phase (see Staubert et al. 2014).

Two features are apparent from Fig.~\ref{fig:history}:
Firstly, we confirm the apparent difference in the mean cyclotron line energy 
before and after 1991, first pointed out by Gruber et al. (2001).
Taking the measured values of E$_{\rm cyc}$ and their stated uncertainties
at face value, the mean cyclotron line energies $\langle E_c \rangle$ 
from all measurements before 1991 is $34.9\pm0.3\,keV$, the corresponding 
value for all measurements between 1991 and 2006 is $40.3\pm0.1\,keV$ 
($40.2\pm0.1\,keV$ for \textsl{RXTE} results only, showing that
the very high value measured by \textsl{BATSE} is not decisive). However, a 
comparison of measurements from different instruments is difficult because of 
systematic uncertainties due to calibration and analysis techniques. 
Nevertheless, we believe that the large difference of $\sim 5\,keV$ 
between the mean values and the good internal consistency within the
two groups (5 different instruments before 1991 and four after 1991) most 
likely indicate real physics.

The first observations with \textsl{RXTE} in 1996 and 1997 showed lower 
E$_{\rm cyc}$ values than those found from \textsl{CGRO}/BATSE and 
\textsl{Beppo}/SAX, leading to the idea of a possible long-term decay.
This idea had then served successfully as an important argument to ask for 
more observations of Her~X-1. In a series of \textsl{RXTE} observations until 
2005 the apparent decrease seemed to continue until this date. At that time
we were determined to publish a paper claiming evidence of a decay 
of the phase averaged cyclotron line energy E$_{\rm cyc}$.
However, working with a uniform set of \textsl{RXTE} data between
1996 and 2005, we discovered that there was a dependence of E$_{\rm cyc}$ 
on X-ray flux (Staubert et al. 2007), degrading the apparent decrease with 
time largely to an artifact: nature seemed to have conspired such that later 
measurements were (on average) taken when the flux happened to be low 
(Her X-1 is known for varying its flux within a factor of two, on timescales of a 
few 35\,d cycles). When the cyclotron line energy was normalized to a common 
flux value, the time dependence almost vanished. 

\begin{figure*}
\includegraphics[width=0.65\textwidth,angle=90]{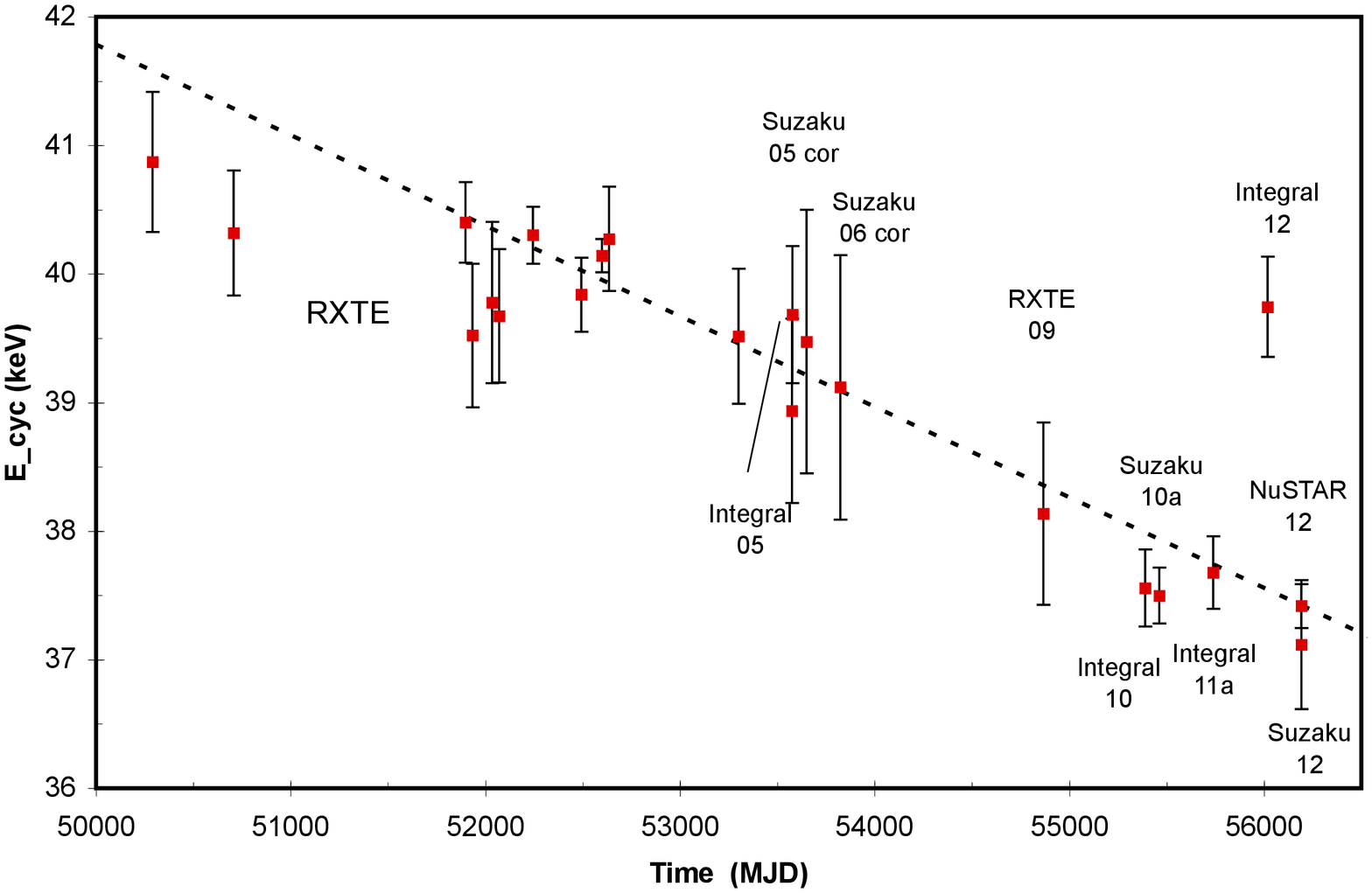}
\hfill
\vspace{-1cm}
\caption{Her~X-1 pulse phase averaged cyclotron line energies E$_{\rm cyc}$ normalized
  to a reference ASM count rate of 6.8\,cts/s using a flux dependence of 0.44\,keV/ASM-cts/s. 
  The data are consistent with a linear decline of E$_{\rm cyc}$ with time 
  with a slope of $-7.22\times 10^{-4}$\,keV${\rm d^{-1}}$ (the dashed line).
 }
 \label{fig:Fnorm3}
\end{figure*}

\subsection{Simultaneous fits to E$_{\rm cyc}$ values with two variables: flux and time}
\label{sec:flux-time_norm3}

With the inclusion of new measurements (2005-2012), we are now able to claim
the first statistically significant evidence of a \textsl{true long-term decay of 
the phase averaged cyclotron line energy} by $\sim 5$\,keV over the last 20 years 
(Staubert et al. 2014). Both dependencies - on flux and on time - seem to 
be always present (they may, however, change their relative importance with time). 
Using a procedure of fitting the E$_{\rm cyc}$ values with two variables simultaneously, 
the two dependencies can be separated and the formal correlation minimized.

We use the following fit function 
\vspace{-1.5mm}
\begin{equation}
{\rm E}_{\rm cyc}(\rm calc) = E_{\rm 0} + a \times (F - F_{0}) + b \times (T - T_{0}) \\
\end{equation}
with F being the X-ray flux (the maximum flux of the respective 35\,d cycle) in units of 
ASM-cts/s, as observed by \textsl{RXTE}/ASM (and/or \textsl{Swift}/BAT), with  
F$_{0}$ = 6.80 ASM-cts/s, and T being time in MJD with T$_{0}$ = 53000.

Staubert et al. (2014) show that the flux dependence is always present, the formal 
value for the slope of the E$_{\rm cyc}$/flux correlation, however, depends slightly on the 
particular data set and fit function used. Assuming no time dependence (b = 0) the data
until 2006 lead to a slope for the flux dependence of $0.58\pm0.1$\,keV/(ASM cts/s).
If this is used to normalize E$_{\rm cyc}$ values of the full data set (1996-2012) to the 
reference flux of 6.8\,(ASM cts/s), the mean $<$E$_{\rm cyc}$$>$ values are: 
$40.1\pm0.1$\,keV for (1996-2006) and $37.6\pm0.1$\,keV for (2007-2012). The difference
of $>17$ standard deviations demonstrates the long-term decrease of E$_{\rm cyc}$.

A fit (flux- and time-variability) to the full data set (1996-2012) with equ.~(4.1) yields
the following best-fit parameters: a = $0.44\pm0.09$\,keV/(ASM cts/s) and 
b = $(-7.22\pm0.39) 10^{-4}$\,keV/d. Dividing b by its uncertainty yields again $>18$ standard
deviations. Fig.~\ref{fig:Fnorm3} shows the evolution in time of the E$_{\rm cyc}$ values
which are normalized to the reference flux of 6.8\,(ASM cts/s) (the dashed line corresponds
to the above fit). The decay of E$_{\rm cyc}$ with time (consistent with a linear decrease) is 
clearly demonstrated.




\vspace{-2mm}
\section{Discussion}
\label{sec:Discussion}
\vspace{-2mm}

\subsection{Dependence of E$_{\rm cyc}$ on luminosity}
\label{sec:on_lum}
\vspace{-2mm}

For a detailed discussion about the observational evidence for both the negative
and the positive correlation of E$_{\rm cyc}$ in different sources, we refer to
Staubert et al. (2014). Here, we just like to mention that Her~X-1 was again the
first source in which the positive correlation (an increase in E$_{\rm cyc}$ with
increasing luminosity) was discovered (Staubert et al. 2007), and that today the
still small group of four objects with a positive correlation now outnumbers
the group of secure sources with the negative correlation discovered earlier.

Our current understanding of the physics behind these correlations assumes that we can
distinguish between \textsl{two accretion regimes} in the accretion column above the polar cap
of the neutron star: \textsl{super- and sub-Eddington accretion}. The former is responsible for
the first detected negative correlation in high luminosity outbursts of transient X-ray sources 
(the reference source being V~0332+53): in this case the deceleration of the accreted 
material is provided by radiation pressure, such that with increasing accretion rate 
$\dot{M}$, the shock and the scattering region move to larger height above the surface of 
the neutron star and consequently to weaker B-field (Burnard et al. 1991). 
Sub-Eddington accretion, on the other hand, leads to the opposite behavior. In this regime 
the deceleration of accreted material is predominantly through Coulomb interactions
and an increase in $\dot{M}$ leads to an increase in electron density (due to an increase
of the combined hydrostatic and dynamical pressure) resulting in a \textsl{squeezing}
of the decelerating plasma layer to smaller height and stronger B-field (Staubert et al. 2007).
More detailed physical considerations have recently been presented by Becker et al. (2012).
The persistent sources Her~X-1 and Vela~X-1are clearly sub-Eddington sources.

\subsection{Dependence of E$_{\rm cyc}$ on time: the long-term variation}
\label{sec:on_time}

With regard to the physical interpretation of the now observed long-term decrease in the 
cyclotron line energy, we speculate that it could be connected to a geometric displacement 
of the cyclotron resonant scattering region in the dipolar field or to a true physical change in 
the magnetic field configuration at the polar cap, which evolves due to continued accretion. 
Apparently, the magnetic field strength at the place of the resonant scattering of photons 
trying to escape from the accretion mound surface has changed with time. 
We suggest that it reflects a local phenomenon in the accretion mound, rather than a change 
in the strength of the underlying global dipole field.

The whole issue of accretion onto highly magnetized neutron stars in binary X-ray sources is 
very complex. A fundamental question for instance is: what happens to the material which is
continuously accreted? Can material be accumulated in
the accretion mound, confined by the B-field? If so, how much, and what effect does this have 
on the field? Or is the material somehow lost at the bottom of the mound - either by leaking to 
larger areas of the neutron star surface or by incorporation into the neutron star crust? Is the 
"gain" and "loss" of material in equilibrium?

We suggest that the observed change of E$_{\rm cyc}$ may be connected to a slight imbalance 
between gain and loss, such that the structure of the column/mound changes. With an accretion 
rate of $\sim 10^{17}$\,g/s a variation on relatively short time scales does not seem implausible.
If the observed decrease in E$_{\rm cyc}$ were due to a simple movement of the resonant 
scattering region to a larger distance from the neutron star surface (possibly caused by a slightly
larger "gain" than "loss"), the observed $\sim$5\,keV reduction in E$_{\rm cyc}$ from 1992 to 2012 
(0.25\,keV per year) would correspond to an increase in height of $\sim$400\,m (for a dipole field).
Alternatively, the configuration of the magnetic field could change with increasing mass in the accretion 
mound: the accreted material could drag the central field lines radially out, thereby diluting the effective 
field strength in the center while enhancing it at larger radii. In modeling magnetic accretion
mounds, Mukherjee \& Bhattachary (2012) have shown that an accumulated mass of 
$\sim 10^{-12}$ $\rm M_{\odot}$ (which is accreted within a few hours) can substantially change 
the field configuration. They also conclude that a small change of mound size could lead to an 
appreciable change in the maximum magnetic field strength. Similar calculations were presented 
by Brown \& Bildsten (1998), Litwin et al. (2001), Payne \& Melatos (2004), 
Payne \& Melatos 2007, asking whether \textsl{screening} or \textsl{burial} of the magnetic field at the 
polar caps is possible by continued accretion, and how much mass could eventually be stored in 
the magnetically confined mounds. While Litwin et al. (2001) work with $\sim 10^{-12}$ $\rm M_{\odot}$
(similar to Mukherjee \& Bhattachary 2012), Payne \& Melatos (2004) need $\sim 10^{-5}$ $\rm M_{\odot}$
to have the magnetic field strength reduced by 10\% through "burying". We note that this amount of mass
seems far to high to be stored in an accretion mound or column. Assume a column with a constant radius
of 1\,km on top of the polar region and a mass of $\sim 10^{-5}$ $\rm M_{\odot}$ piled up in this column: the height 
of the column is then $\sim 10^{5}$\,km, assuming a constant density comparable to that of the neuron star crust 
($10^{8}$\,g/cm$^{3}$). This is clearly unphysical!

In principle, the observed effect of the varying cyclotron line energy could also be due to more exotic effects in the 
accretion mound or in the neutron star crust. Some of the ideas that can be found in the literature, like Ohmic dissipation 
(happening on the magnetic diffusion time scale) or hydrodynamic flows, are mentioned in Staubert et al. (2014).

We finally speculate on a possible cyclic behavior of E$_{\rm cyc}$ on timescales
of a few tens to hundreds of years. Could it be that the fast rise of the observed E$_{\rm cyc}$
value after 1991 (see Fig.~\ref{fig:history}) represents a special event in which the magnetic
field in the accretion mound has rearranged itself as a result of a sudden radial
outflow of material? In the above mentioned model calculations
the field configuration is shown to change considerably with increased material, leading to a 
\textsl{ballooning} of the field configuration with diluted field in the symmetry center and increased 
density of field lines at the circumference of the mound. 
It remains unclear, how important
continuous leaking through the outer magnetic boundary may be and what the timescales for 
semi-catastrophic events might be, in which the field would release (on a short timescale) a substantial 
fraction of the stored material to larger areas of the neutron star surface. For Her~X-1, this scenario could 
mean that we are now in a phase of continuous build-up of the accretion mound with
the mass (and the height?) of the mound growing and the observed cyclotron line energy 
continuously decreasing until another event like the the one around 1991 happens again. The
mean E$_{\rm cyc}$ value measured before 1991 of $\sim$35\,keV may represent a bottom
value. So, when the current decay continues steadily, one may expect another event of
a rather fast increase in E$_{\rm cyc}$ within, say, another 10 to 20 years.

We like to urge both observers and model builders to continue to accumulate
more observational data as well as more understanding of the physics responsible for
the various observed properties in Her~X-1 and other objects of similar nature. 
Our hope is that the new observational result may boost the motivation for further 
theoretical studies. For model builders a challenge would be to work towards
dynamical computations that might eventually lead to self-consistent solutions of the structure and 
evolution of magnetized accretion mounds of accreting neutron stars with only a few input parameters. 



\end{document}